\documentclass[aps,prd,showpacs,nofootinbib,twocolumn,10pt]{revtex4-1}
\usepackage{color,amsmath,amssymb,graphicx,latexsym}
\usepackage{threeparttable,multirow,txfonts,subfigure,booktabs}

\begin{document}

\title{Explaining the GeV antiproton/
$\gamma-$ray excesses and W-boson mass anomaly in an inert two Higgs doublet model}
\author{Cheng-Rui Zhu$^{1,2}$, Ming-Yang Cui$^{1*}$, Zi-Qing Xia$^{1}$, 
Zhao-Huan Yu$^3$, Xiaoyuan Huang$^{1,4*}$,
Qiang Yuan$^{1,4*}$, Yi-Zhong Fan$^{1,4*}$}
\address{$^1$Key Laboratory of Dark Matter and Space Astronomy, Purple
Mountain Observatory, Chinese Academy of Sciences, Nanjing 210023, China\\
$^2$Shandong Institute of Advanced Technology, Jinan 250100, China\\
$^3$School of Physics, Sun Yat-Sen University, Guangzhou 510275, China\\
$^4$School of Astronomy and Space Science, University of Science and
Technology of China, Hefei 230026, Anhui, China
}

\email{mycui@pmo.ac.cn (MYC)}
\email{xyhuang@pmo.ac.cn (XH)}
\email{yuanq@pmo.ac.cn (QY)}
\email{yzfan@pmo.ac.cn (YZF)}

\begin{abstract}
For the newly discovered $W$-boson mass anomaly, one of the simplest dark matter 
(DM) models that can account for the anomaly without violating other astrophysical/experimental 
constraints is the inert two Higgs doublet model, in which the DM mass ($m_{S}$) is found to be 
within $\sim 54-74$ GeV. In this model, the annihilation of DM via $SS\to b\bar{b}$ 
and $SS\to WW^{*}$ would produce antiprotons and gamma rays, and may account for the excesses 
identified previously in both particles. Motivated by this, we re-analyze the AMS-02 antiproton 
and Fermi-LAT Galactic center gamma-ray data. For the antiproton analysis, the novel treatment 
is the inclusion of the charge-sign-dependent three-dimensional solar modulation model as 
constrained by the time-dependent proton data. We find that the excess of antiprotons is more 
distinct than previous results based on the force-field solar modulation model. The interpretation 
of this excess as the annihilation of $SS\to WW^{*}$ 
($SS\to b\bar{b}$) 
requires a DM mass of 
$\sim 40-80$ ($40-60$) 
GeV and a velocity-averaged cross section of $O(10^{-26})~{\rm cm^3~s^{-1}}$. 
As for the $\gamma$-ray data analysis, besides adopting the widely-used spatial template 
fitting, we employ an orthogonal approach with a data-driven spectral template analysis. 
The fitting to the GeV $\gamma$-ray excess yields DM model parameters overlapped with those to 
fit the antiproton excess via the $WW^{*}$ channel. The consistency of the DM particle properties required to account 
for the $W$-boson mass anomaly, the GeV antiproton excess, and the GeV $\gamma$-ray excess
suggest a common origin of them.
\end{abstract}

\maketitle
 
\section{Introduction}

The dark matter (DM) problem remains one of the biggest mysteries of the cosmos.
Among many kinds of candidates, the weakly interacting massive particle (WIMP)
is the most naturally motivated by the thermal production in the early Universe
and its proper relic density today \cite{2005PhR...405..279B}. 
Quite a lot of efforts have been spent in looking for WIMP DM in various kinds 
of experiments. No convincing signal has been identified in the direct detection 
experiments and very stringent constraints on the WIMP-nucleon interaction strength 
have been set (e.g., \cite{Liu:2017drf,Schumann:2019eaa}). 
As for the indirect detection aiming to identify the products of the annihilation 
or decay of the DM particles \cite{1997NuPhB.504...27B,2005PhR...405..279B}, 
some anomalies have been claimed in the past decade, such as the positron and 
electron excesses \cite{PAMELA:2008gwm,AMS:2021nhj,DAMPE:2017fbg}, the antiproton excess\footnote{See also Refs. \cite{Moskalenko:2002yx,Hooper_2015} 
for possible hints of excess from measurements prior to AMS-02.} \cite{Cui_2017,Cuoco_2017}, and the Galactic 
center $\gamma$-ray excess (GCE; \cite{Hooper:2010mq,Zhou:2014lva,Daylan:2014rsa,Calore:2014xka}). 
While the positron and electron excesses might be naturally explained by
astrophysical pulsars \cite{Hooper:2008kg,Yin:2013vaa} and the DM interpretation 
is severely constrained by $\gamma$-ray and cosmic microwave background observations
\cite{Yuan:2017ysv}, the antiproton excess and the GCE which point to a consistent DM 
interpretation survive other constraints \cite{Cui:2018klo,Cui:2018nlm,Cholis:2019ejx}.
In spite that uncertainties of various astrophysical and particle physics ingredients exist 
\cite{Lin:2016ezz,Jin:2017iwg,Heisig:2020nse,Jueid:2022qjg,Bartels:2015aea,Lee:2015fea,Macias:2016nev,Bartels:2017vsx,Leane:2019xiy,Zhong:2019ycb,Cholis:2021rpp}, common implications on the DM scenario from multi-messengers are very interesting. 
In any case, additional tests of this scenario from independent probes are very
important in finally detecting DM particles.

Very recently, the measured $W$-boson mass by the CDF collaboration shows $\sim7\sigma$
deviation from the prediction of the standard model (SM), which strongly suggests the 
existence of new physics beyond the SM \cite{CDF2022}. One of the most economic solutions 
is to introduce an additional scalar doublet, in which the non-SM scalars can enhance 
the $W$-boson mass via the loop corrections. With a proper discrete symmetry $\mathbb{Z}_2$, 
the lightest new scalar in the doublet can be stable and play the role of DM. One realization 
of this mechanism is the inert two Higgs doublet model (i2HDM), which is shown to be able
to accommodate the new $W$-boson mass and various astrophysical/experimental constraints 
simultaneously \cite{i2HDM}. Considering available constraints from the collider searches
for new physics, the electroweak precision tests, the direct detection of DM, and the
relic density of DM, the mass of DM is limited within the range of 
$54~{\rm GeV}<m_S<74~{\rm GeV}$, and the annihilation is dominated by the process of 
$SS\rightarrow WW^{*}$ for $m_S\geq 62$ GeV and by $SS\rightarrow b\bar{b}$ otherwise. 

It is thus essential to examine whether the astrophysical data are in support of such 
an attractive possibility or not. For such a purpose, we re-analyze the AMS-02 antiproton
and Fermi-LAT Galactic center $\gamma$-ray data. Compared with previous works, we improve
the technical treatments in several aspects to reduce potential uncertainties of the
analyses. For the antiproton modeling, our novel treatment is to include the 
charge-sign-dependent three-dimensional (3D) solar modulation model
\cite{2015ApJ...810..141P,2019ApJ...878....6L} as constrained by the time-dependent AMS-02 proton data \cite{AMS:2018qxq}. 
{To investigate the GCE, taking a data-driven method, we identify the background components for the $\gamma$-ray sky solely with their spectral  properties as in Ref.~\cite{Huang:2015rlu}, called as the spectral template analysis. The traditional spatial template analysis will also be employed for a cross check.}
To minimize the possible contamination from the astrophysical contribution in the galactic bulge \cite{Macias:2016nev,Bartels:2017vsx}, a large portion of the galactic disk is masked. 
We find consistent DM particle properties to account for the $W$-boson mass anomaly, the GCE, and the antiproton excess, which are in favor of a common origin.

\section{Antiprotons}\label{Antiprotons}
In previous studies, the solar modulation of the antiprotons is usually assumed to be 
the same as that of protons and the force-field approximation \cite{1968ApJ...154.1011G} 
was often adopted\footnote{An empirical approach to derive a time, rigidity, and 
charge-sign dependent force-field modulation potential has been developed in 
Ref.~\cite{Cholis:2015gna}.}. However, it is known that the particles with opposite
charge have very different trajectories in the heliosphere (e.g., \cite{2013PhRvL.110h1101M}). 
Such an effect should be taken into account to properly reproduce the local interstellar 
spectra (LIS) of protons and antiprotons. For such a purpose, here we employ the 
charge-sign-dependent 3D solar modulation model developed in 
\cite{1999ApJ...513..409Z,2011ApJ...735...83S}. The transportation of charged 
particles inside the heliosphere is described by the Parker's equation 
\cite{1965P&SS...13....9P}
\begin{equation}\label{eq:parker}
    \begin{split}
    \frac{\partial f}{\partial t} =  & -\left( \boldsymbol{V_{\rm sw}}+ \left\langle\boldsymbol{v_d}\right\rangle \right) \cdot \nabla f+ \nabla \cdot \left(\boldsymbol{K^{(s)}} \cdot \nabla f\right)\\
    &+\frac{1}{3} \left(\nabla \cdot \boldsymbol{V_{\rm sw}}\right) 
    \frac{\partial f}{ \partial \ln p},
\end{split}
\end{equation}
where $f(\boldsymbol{r},p,t)$ is the phase space distribution function of cosmic rays, 
$\boldsymbol{V_{\rm sw}}$ is the solar wind speed, $\left\langle\boldsymbol{v_d}\right\rangle$ 
is the pitch-angle-averaged drift velocity, $\boldsymbol{K^{(s)}}$ is the symmetric 
diffusion tensor, and $p$ is the momentum of the particle. See the Appendix Sec. A 
for more details. We solve the Parker's equation numerically employing the stochastic 
differential equations \cite{1999ApJ...513..409Z,2011ApJ...735...83S}. 

\begin{figure}[!ht]
\centering
\includegraphics[width=\columnwidth]{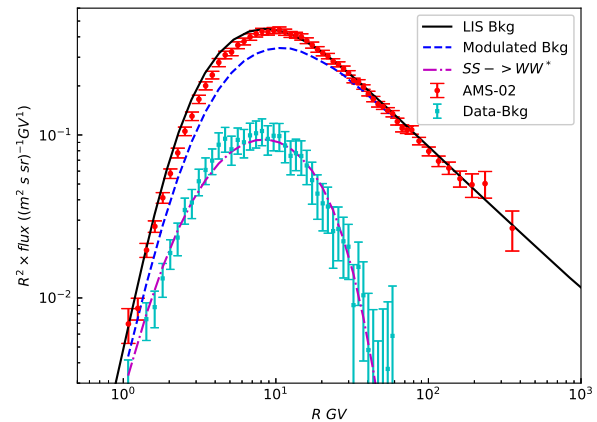}  
\caption{The LIS (solid) and the modulated (dashed) spectra of background antiprotons, 
compared with the data \cite{AMS:2016oqu}. The cyan points denote the AMS-02 
data minus the modulated background results, and the dash-dotted line shows the antiproton 
spectrum from DM annihilation into $WW^*$ with ($m_S$, $\left\langle \sigma v \right\rangle$) 
= (62.6 GeV, $1.5 \times 10^{-26}~{\rm cm}^3~{\rm s}^{-1}$).} 
\label{fig:ap}
\end{figure}

The LIS of protons is derived through fitting to the Voyager-1 \cite{2016ApJ...831...18C}, 
AMS-02 \cite{AMS:2021nhj}, and DAMPE \cite{DAMPE:2019gys} data. To do this fitting,
we employ the {\tt GALPROP}\footnote{\href{https://galprop.stanford.edu/}
{https://galprop.stanford.edu/}} code to calculate the propagation of cosmic rays in 
the Milky Way \cite{1998ApJ...509..212S}. The detailed fitting procedure is described
in the Appendix Sec. B. The antiproton LIS, calculated based on the proton LIS, is shown 
by the black solid line in Fig.~\ref{fig:ap}. Here we use the new parameterization 
of the antiproton production cross section from Ref.~\cite{Winkler:2017xor} and an 
energy-dependent nuclear enhancement factor to take into account the contribution from 
heavy nuclei in both cosmic rays and the interstellar medium \cite{Kachelriess:2015wpa}.

We fit the time-dependent proton fluxes measured by AMS-02 
\cite{AMS:2018qxq} to obtain the solar modulation parameters. The AMS-02
monthly proton fluxes are grouped into 9 time bins, each contains 6 Bartels rotations, 
from May 19, 2011 to May 26, 2015, corresponding to the antiproton measurement time
\cite{AMS:2016oqu}. The fitting results of the main modulation parameters 
are given in the Appendix Sec. C. Using the best-fit parameters, we calculate the 
modulated antiproton spectrum, as shown by the blue dashed line in Fig.~\ref{fig:ap}.
We find that, the modulated background spectrum from the cosmic ray interactions
is lower than the data between 1 and 30 GeV, consistent with previous studies
\cite{Cui_2017,Cuoco_2017,Cui:2018klo}. Intriguingly, the difference between the 
antiproton data and the predicted astrophysical background is more distinct than that 
found previously with the force-field solar modulation. This is perhaps due to that
particles with negative charge were modulated more severely than positive charged
particles after the reversal of the heliospheric magnetic field \cite{Zhu:2020koq}.

Then we consider the DM contribution to the antiprotons. The DM density distribution
is assumed to be a generalized Navarro-Frenk-White profile \cite{1996MNRAS.278..488Z}, 
with a local density of 0.4 GeV cm$^{-3}$ and an inner slope of 1.28\footnote{This
density profile is consistent with the prediction from the cold DM model, which is largely
consistent with observations of DM dominated systems, such as Ultra Faint Dwarf galaxies
\cite{Simon:2019nxf,Hayashi:2020syu}.}. The annihilation 
into $b\bar{b}$ or $W^+W^-$ is considered. For DM mass $m_S<m_W$, we also consider 
the off-shell annihilation into $WW^*$, as in the case of i2HDM. The DM annihilation 
into $WW^*$ is simulated with \texttt{MadGraph5\_aMC@NLO} \cite{Alwall:2014hca}, including 
all three-body final states of one on-shell and one off-shell $W$ bosons. We further 
utilize \texttt{PYTHIA 8} \cite{Sjostrand:2014zea} to carry out the simulation of final 
state radiation, hadronization, and particle decays, and obtain the corresponding energy 
spectra of antiprotons and $\gamma$ rays. 

We perform a likelihood fitting to the antiproton data, with a marginalization of the 
constant re-scaling factor of the background, and obtain the constraints on the 
$(m_S,\langle\sigma v\rangle)$ parameters. The results are shown in Fig.~\ref{fig:msv} 
{for the $WW^*$ channel, and the results for the $b\bar{b}$ channel are shown in the Appendix Sec. G}. 
The favored mass of DM particles is from $40$ to $60$ GeV for the $b\bar{b}$ channel, 
and from $40$ to $80$ GeV for the $WW^{*}$ channel, respectively, and the annihilation 
cross section is around the level of the thermal production of DM, i.e., 
${\cal O}(10^{-26}~{\rm cm^{3}~s^{-1}})$.
{We also consider the influence of uncertainties of solar modulation on the likelihood fitting. The result is shown in the Appendix Sec. D}.
We can see that the contours of $WW^*$ overlap 
well with the i2HDM model parameters to fit the $m_W$ anomaly \cite{i2HDM}. Two possible 
regions of the i2HDM parameter space with DM mass of about $70-73$ GeV (for four-point 
interactions) and about $62-63$ GeV (for the Higgs resonance and scalar-pseudoscalar 
co-annihilation region) can commonly account for the antiproton excess and the $m_W$ anomaly. 
In the even lower-mass window ($m_S<62$ GeV), the i2HDM model to fit the $m_W$ anomaly 
typically requires DM to annihilate dominantly into $b\bar{b}$ but with a much smaller 
cross section and seems not able to produce enough antiprotons. 

\begin{figure}[!ht]
\centering
\includegraphics[width=\columnwidth]{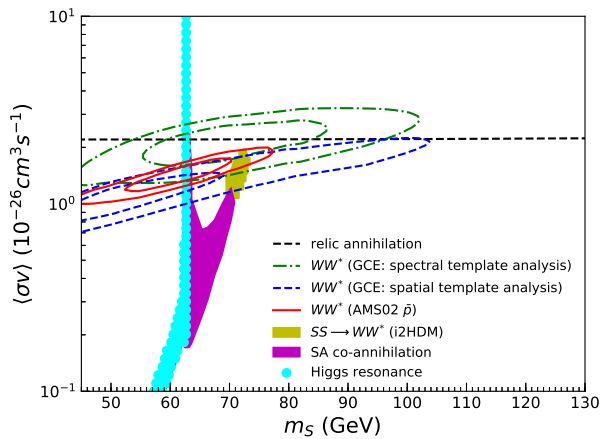}  
\caption{The favored DM parameter space via fitting to the antiproton and GCE data (1$\sigma$ and 2$\sigma$ from inside to outside) for the $WW^*$ channel, as well as the i2HDM model parameters to fit the $W$-boson mass anomaly (the 95\% region, adopted from Ref.~\cite{i2HDM}).
The black dashed line is the mass-dependent relic annihilation cross section \cite{Steigman2012}.} \label{fig:msv}
\end{figure}

\section{Galactic center gamma rays}

The Galactic center is expected to gather high densities of DM, which makes it appealing for the indirect detection of DM. A spatially extended excess of $\gamma$ rays in the $1–10$ GeV energy range, with respect to the expected emission of cataloged point sources and astrophysical diffuse components, was found in the Fermi-LAT observations (e.g.,
\cite{Hooper:2010mq,Zhou:2014lva,Calore:2014xka,Daylan:2014rsa}), named as the GCE. 

{To investigate this GeV excess, Refs.~\cite{Selig2014,Huang:2015rlu,deBoer:2016esu, deBoer:2017sxb} adopt the spectral template analysis, which attempts to reduce the influences from uncertainty in spatial templates of astrophysical diffuse components.
The $\gamma$-ray sky is decomposed into point sources and diffuse emission component by D$^{3}$PO and spectra from the ``cloud-like'' component and the ``bubble-like'' component, for diffuse emission from hadronic and leptonic processes, were derived from two distinctive regions \cite{Selig2014}. 
The astrophysical background is investigated by including point sources and data-driven spectral templates. 
Further, the spectral information of DM annihilation is used as a new component. The existence of the GCE was confirmed and consistent DM parameters were derived in this alternative way if the spatial distribution of the GCE is fixed as in Ref.~\cite{Calore:2014xka}, although an indication of a spatial correlation between the GCE and astrophysical component was shown if the spatial distribution of the GCE is free \cite{Huang:2015rlu}. Note that following Ref.~\cite{Huang:2015rlu}, we also consider the isotropic component in the spectral template analysis. Though this component is sub-dominant in the Galactic center region, it would be important in the region far away from the galactic disk. See more details of the spectral template analysis in Ref.~\cite{Huang:2015rlu}.} 

Besides the annihilation channel of $SS\to b\bar{b}$~\cite{Huang:2015rlu}, here we apply this method to investigate whether the annihilation of DM via $SS\rightarrow WW^{*}$ could be consistent with the Galactic center $\gamma$-ray data. 
The $\gamma$-ray dataset, the point sources, and the spectra from the “cloud-like” component, the “bubble-like” component {and the isotropic $\gamma$-ray component} are the same as in Ref.~\cite{Huang:2015rlu}. Here the DM density distribution is 
the same as that for the antiproton analysis in Sec.~\ref{Antiprotons}. The region of interest 
(ROI) used in this analysis is a square region selected by Galactic latitudes $\left| b\right| \textless 20^{\circ}$ and Galactic longitudes $\left| l\right| \textless 20^{\circ} $, with a mask of the Galactic plane with $\left| b\right| \textless 8^{\circ}$, 
to eliminate the influence from the boxy bulge, the nuclear bulge and the X-shaped bulge~\cite{Macias:2016nev,Bartels:2017vsx}.

We scan the DM parameters to calculate 
the likelihood map of different parameters. See the Sec. E of the Appendix for the selected 
ROI and likelihood map of the GCE fitting. 
{The best fit DM parameters we got are ($m_S$, $\left\langle \sigma v \right\rangle$) = (68.5 GeV, $2.1 \times 10^{-26}~{\rm cm}^3~{\rm s}^{-1}$) for the $WW^*$ channel,  and (77.5 GeV, $1.7 \times 10^{-26}~{\rm cm}^3~{\rm s}^{-1}$) for the $ b\bar{b}$ channel.}
{The 1$\sigma$ and 2$\sigma$ contours of the fittings ($WW^*$) are given in Fig.~\ref{fig:msv} by green dash-dotted lines.}
At $2\sigma$ level, the favored region of the $WW^*$ channel from the GCE, the antiproton excess, and the $m_W$ anomaly overlap with each other. 
Given that there should be additional uncertainties from various aspects of the theoretical modeling (e.g., the density profile of DM in the inner Galaxy and the simulation of the spectra of the annihilation final state particles), 
we regard these three anomalies are accounted for simultaneously with the same DM model component.
{The results for the $b\bar{b}$ channel are shown in the Appendix Sec. G.}

{As a complementary check, we also take the traditional spatial template regression techniques~\cite{Zhou:2014lva, Calore:2014xka, Cholis:2021rpp} to fit the Fermi-LAT $\gamma$-ray observations to investigate the GCE. See the Appendix Sec. F for more details of the spatial template analysis.
Corresponding to the minimums of mean $\chi^2$ values, the best fit DM parameters ($m_S$, $\left\langle \sigma v \right\rangle$) we obtained are (49.7 GeV, $1.0 \times 10^{-26}~{\rm cm}^3~{\rm s}^{-1}$) for the $WW^*$ channel and (50.1 GeV, $6.7 \times 10^{-27}~{\rm cm}^3~{\rm s}^{-1}$) for the $b\bar{b}$ channel, respectively.
The 1$\sigma$ and 2$\sigma$ contours ($WW^*$) for this spatial template analysis are shown in Fig.~\ref{fig:msv} as blue dashed lines. 
Same as the result of the previous spectral template analysis, we also find the $2\sigma$ confidence region of the $WW^*$ channel from the GCE, the antiproton excess and the $m_W$ anomaly invariably overlap with each other.}

\section{Conclusion and discussion}
Very recently, the CDF collaboration reported a statistically significant $W$-boson mass excess \cite{CDF2022}, which strongly indicates the new physics beyond the SM \cite{Lu:2022bgw,Athron:2022qpo}. One interesting possible solution is the i2HDM, which  indicates dark matter particles with a mass of $\sim 50-70$ GeV, and with the cross section $\sim {\cal O}(10^{-26}~{\rm cm^{3}~s^{-1}})$ for the $SS\rightarrow WW^*$ annihilation channel. If correct it might yield observable gamma-rays and anti-protons in the Galaxy. Motivated by such a possibility, in this work we re-analyze the Fermi-LAT gamma-ray and AMS-02 anti-proton data and then investigate the possible DM origin of the identified excesses. While our 
excess signals are generally in agreement with previous works, we incorporate several new technical treatments in the analysis such as the charge-sign-dependent 3D solar modulation of anti-protons and a spectral template fitting scheme of $\gamma$ rays, as well as the
off-shell annihilation channel of $SS\to WW^*$. It is very intriguing to find that the three very different anomalies can be simultaneously accounted for in a minimal DM model with DM particle mass of $\sim 60-70$ GeV. The velocity-weighted annihilation cross section is about $\left\langle\sigma v \right\rangle\sim O(10^{-26}~{\rm cm^{3}~s^{-1}})$ and is just consistent with the expectation of the thermal production of DM. {The required DM parameters are also consistent with constraints from other probes such as neutrinos \cite{IceCube:2021kuw}.}
Although there are various kinds of uncertainties of the antiproton background calculation like the propagation model of cosmic rays, the hadronic interaction models, and/or the solar modulation which is partially addressed in this work \cite{Lin:2016ezz,Jin:2017iwg,Heisig:2020nse,Jueid:2022qjg}, as well as debates of the astrophysical or DM origin of the GCE \cite{Bartels:2015aea,Lee:2015fea,Macias:2016nev,Bartels:2017vsx,Leane:2019xiy,Zhong:2019ycb,Cholis:2021rpp},
the DM interpretation of the three independent signals seems to be a straightforward, economic, and attractive possibility. The ongoing direct detection experiments such as 
the PandaX-4T, Xenon-nT and LUX has a good prospect to detect it in the near future, 
as shown in \cite{i2HDM}. 

{We comment that the antiproton excess identified in $1-40$ GeV (Fig.~\ref{fig:ap}) is likely hard to be accounted for by astrophysical sources. As a commonly proposed scenario that secondary particles may also be produced via interactions around the accelerating sources (e.g., supernova explosion in molecular clouds), harder spectra of secondary particles are expected which should be more evident at high energies \cite{Biermann:2018clk,Zhang:2021xri}. If we artificially attribute the identified low-energy antiproton excess to an astrophysical secondary particle component, the corresponding B/C ratio would be significantly higher than the measurements, as shown in the Supplemental Material of Ref.~\cite{Cui_2017}.}

\acknowledgments
This work is supported by the National Key R$\&$D program of China (No. 2021YFA0718404), the National Natural Science Foundation of China (No. 11921003, 11903084, 12003069, 12220101003), 
Chinese Academy of Sciences (CAS), 
the CAS Project for Young Scientists in Basic Research (No. YSBR-061), 
and the Program for Innovative Talents and Entrepreneur in Jiangsu. 

\setcounter{figure}{0}
\renewcommand\thefigure{A\arabic{figure}}

\section*{Appendix}
\subsection{Setup of the 3D modulation model}

We describe in more detail the ingredients of the 3D modulation model. 
The diffusion tensor, $\boldsymbol{K^{(s)}}$, given in the heliospheric 
magnetic field (HMF) aligned coordinates, is
\begin{equation}
    \boldsymbol{K^{(s)}} = \left( \begin{array}{ccc} \kappa_{\parallel} &0 &0\\0& 
    \kappa_{\perp \theta}& 0\\ 0& 0& \kappa_{\perp r}\\
    \end{array} \right),
\end{equation}
where $\kappa_{\parallel}$ is the parallel diffusion coefficient, $\kappa_{\perp r}$ 
and $\kappa_{\perp\theta}$ are the two perpendicular diffusion coefficients in the 
radial and latitudinal directions. In our work, we parameterize the parallel diffusion 
coefficient as \cite{2015ApJ...810..141P,2017ApJ...839...53L}
\begin{equation}
\begin{split}
    \kappa_{\parallel} = \kappa_{\parallel}^0 \beta\frac{B_0}{B}\left(\frac{p}{p_0}\right)^{\alpha_1} 
    \left[\frac{\left(\frac{p}{p_0}\right)^s + \left (\frac{p_k}{p_0}\right)^s}
    {1+\left(\frac{p_k}{p_0}\right)^{s}}\right]^{\frac{\alpha_2-\alpha_1}{s}},
\end{split}
\end{equation}
where $\beta$ is the particle speed in unit of the speed of light, $B$ is the magnitude 
of the HMF with $B_0= 1 ~{\rm nT}$, $p$ is the particle momentum, $p_0=1~{\rm GV}$,
and $s$ is the smoothness transition parameter which is fixed to be 2.2. The perpendicular 
diffusion coefficients are assumed to be $\kappa_{\perp r,\perp \theta}=0.02\kappa_{\parallel}$. 
The free parameters of the diffusion coefficient are $\kappa_{\parallel}^0$, $\alpha_1$, 
$\alpha_2$, and $p_k$.

We adopt the standard Parker HMF model \cite{1958ApJ...128..664P} as
\begin{equation}
    \boldsymbol{B}(r,\theta)=\frac{A_0B_\oplus}{r^2}\left(\boldsymbol{e_r} - \frac{r\Omega \sin \theta}{V_{\rm sm}} \boldsymbol {e_\theta}\right)\left[1-2H(\theta-\theta_{\rm cs})\right],
\end{equation}
where $B_\oplus$ is the reference value at the Earth's location, 
$A_0 = A\sqrt {1+\Omega^2/V_{\rm sw}^2}$ with $A = \pm 1$ describing the polarity 
of the HMF, $\Omega$ is the angular velocity of the Sun, $H(\theta-\theta_{\rm cs})$ 
is the Heaviside function and $\theta_{\rm cs}$ determines the polar extent of the 
Heliospheric Current Sheet (HCS). Following Ref.~\cite{1983ApJ...265..573K}, we have
\begin{equation}
    \cot(\theta_{\rm cs})=-\tan (\alpha) \sin(\phi^*),
\end{equation}
where $\alpha$ is the HCS tilt angle, $\phi^*=\phi+r\Omega/V_{\rm sw}$ is the foot 
point at the Sun for the corresponding spiral magnetic field line with $\phi$ 
being the longitude angle of the current sheet surface. 

The drift velocity $\left \langle \boldsymbol{v_d} \right \rangle$ can be described as \cite{1977ApJ...213..861J,1989ApJ...339..501B}  
\begin{equation}\label{eq:drift}
\left\langle\boldsymbol{v_d}\right\rangle =\frac{pv}{3qB} \nabla \times \frac{\boldsymbol{B}}{B},
\end{equation}
where $q$ is the particle's charge and $v$ is the velocity. To avoid the singularity
of the drift close to the HCS where the direction of the magnetic field changes abruptly, 
the drift velocity within $(-2R_{\rm g}, 2R_{\rm g})$ to the HCS is approximated as $v/6$ 
in magnitude and the direction is along the HCS \cite{1989ApJ...339..501B,2017ApJ...839...53L}. 
Here $R_{\rm g}$ is the gyro radius of the particle.
As discussed in Ref.~\cite{1989JGR....94.2323P}, the drift velocity may be suppressed, 
especially for when the solar activity is strong. Here we multiply a factor 
$K_{\rm d} \in [0,1]$ on the drift velocity and fit $K_d$ with the data.

\subsection{The proton LIS}

\begin{figure}[!htb]
\centering
\includegraphics[width=\columnwidth]{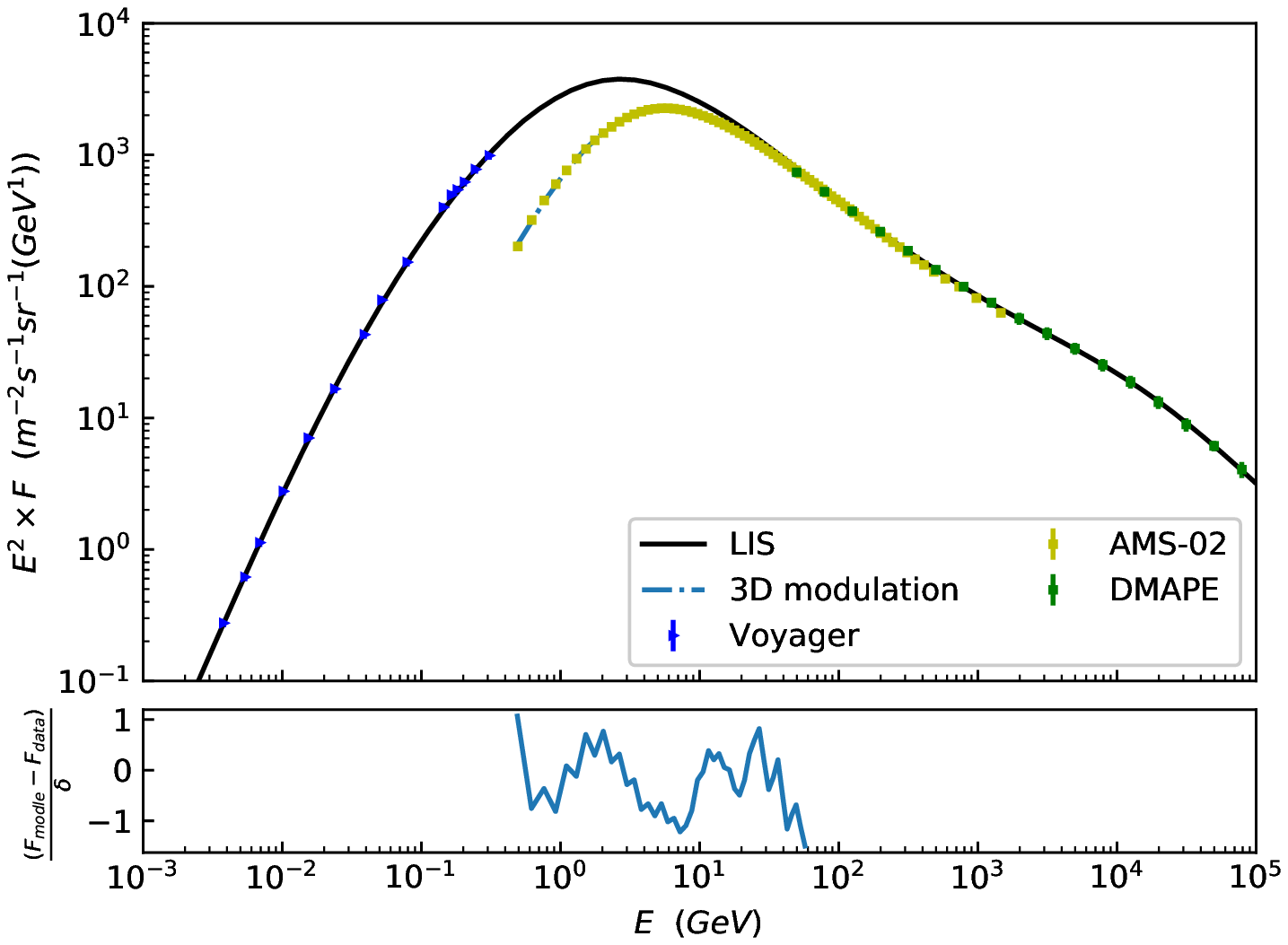}  
\caption{The LIS (solid) and modulated ( dash-dotted
for the 3D modulation) spectrum of protons, compared with the Voyager-1
\cite{2016ApJ...831...18C}, AMS-02 \cite{2015PhRvL.114q1103A}, and DAMPE \cite{DAMPE:2019gys} data. The lower sub-panel  shows the residuals for 3D modulation. 
} 
\label{fig:plis}
\end{figure}

The Voyager-1 measurements of proton fluxes outside the heliosphere \cite{2016ApJ...831...18C} 
and the AMS-02 measurements at high energies (e.g., $\gtrsim 50$ GeV) \cite{AMS:2021nhj} 
can robustly constrain the low- and high-energy parts of the proton LIS. However, for the 
intermediate energy range, the LIS entangles with the solar modulation, and thus a 
model-dependence is unavoidable. The antiproton LIS depends further on the distribution and 
interactions of protons in the Milky Way as well as the interstellar propagation of particles. 
The {\tt GALPROP} code is used to calculate the propagation of cosmic rays. The propagation 
parameters were derived through fitting the secondary and primary nuclei of cosmic rays 
\cite{yuan2020}. The reacceleration model is adopted as the benchmark of this work since it 
fits the secondary nuclei data much better than the model without reacceleration. We use a 
cubic spline method \cite{zhucr2018} to describe the source spectrum of protons, and employ 
the force-field modulation model to approximate the average solar modulation of protons during 
the data-taking period of AMS-02 \cite{AMS:2021nhj}. The solid line in Fig.~\ref{fig:plis} 
shows the calculated LIS of protons using the best-fit source parameters. 
Also shown are AMS-02 proton date from \cite{2015PhRvL.114q1103A} with the data-taking period 
is coverd by the monthly AMS-02 proton data, and the 3D modulation (dash-dotted) with parameters 
derived in this work (see below). The differences between the 3D modulated spectrum and data 
is very small.

\subsection{Fit to the time-dependent proton fluxes}

\begin{figure}[!ht]
\centering
\includegraphics[width=\columnwidth]{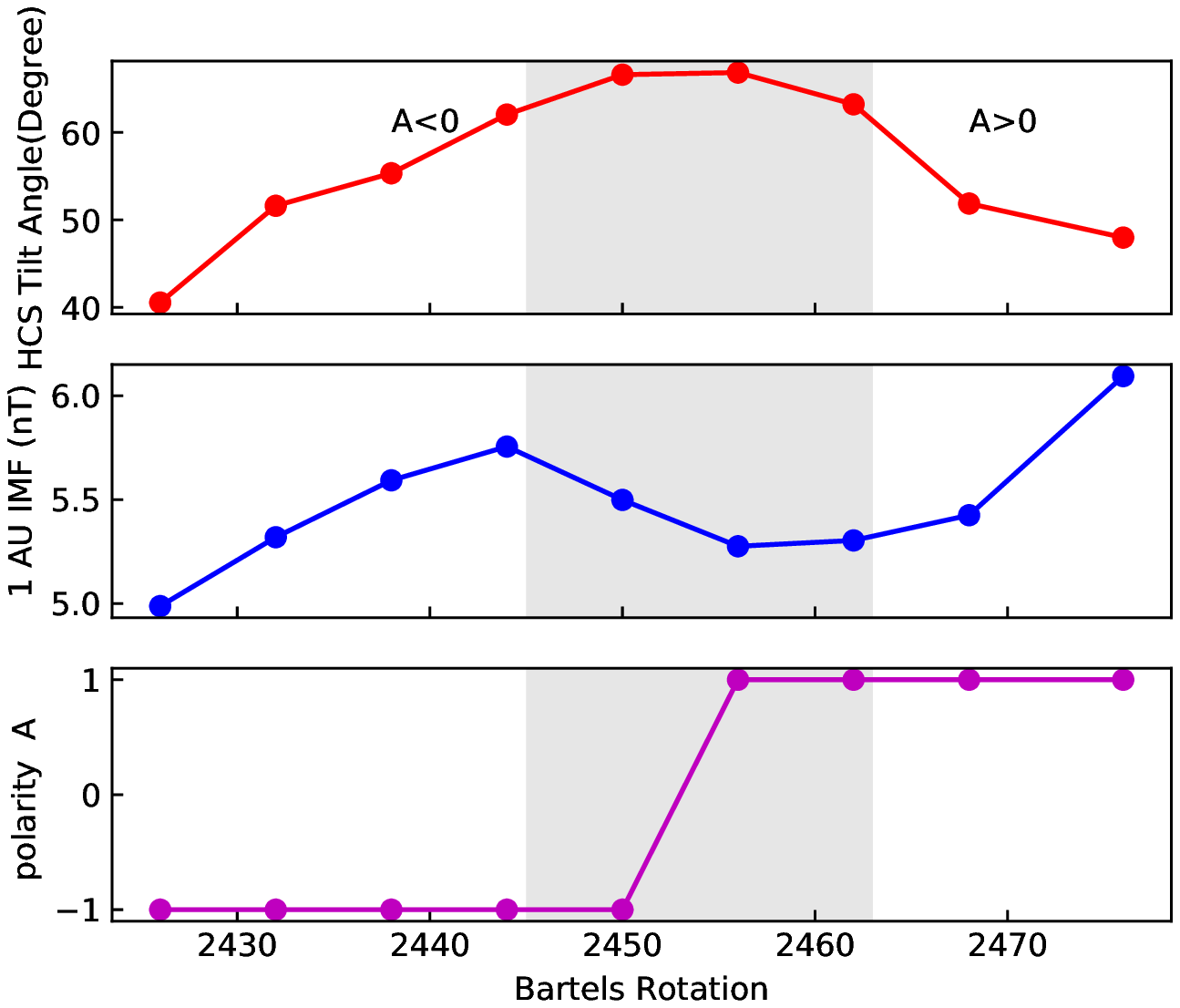}  
\caption{Top: the HCS tilt angle from the WSO. Middle: the average HMF values from the ACE.
Bottom: the polarity of the HMF adopted in this work. The gray belts denote the period 
when the polarity is uncertain \cite{2015ApJ...798..114S}.} 
\label{fig:MF_TA_A}
\end{figure}

The monthly proton fluxes reported in \cite{AMS:2018qxq} are used to derive
the time-dependent solar modulation parameters. To reduce the computation load, the monthly 
data are rebinned into nine time bins with Bartels rotations: $2426 - 2431$, $2432 -2437$, 
$2438 - 2443$, $2444 - 2449$, $2450 - 2455$, $2456 -2461$, $2462 - 2467$, $2468 - 2475$, 
and $2476 - 2479$, which cover the time interval of the antiproton measurements \cite{AMS:2016oqu}. Note that there is no data during Bartels rotations $2472 - 2473$.
Since there is about one year time lag between the cosmic ray flux modulation and solar 
activities \cite{zhucr2018}, following Ref.~\cite{2019ApJ...878....6L}, we utilize the 
previous 13 month averaged interplanetary parameters for every month, 
including the HCS tilt angle from the Wilcox Solar Observatory (WSO\footnote{\href{http://wso.stanford.edu}{http://wso.stanford.edu}}), 
the average HMF values from the Advanced Composition Explorer 
(ACE\footnote{\href{http://www.srl.caltech.edu/ACE/ASC/index.html}
{http://www.srl.caltech.edu/ACE/ASC/index.html}}). 
As the time bin we use here is about half a year, we further take the six month mean 
values of the interplanetary parameters. As for the polarity, we assume $A<0$ before 
Bartels rotation 2455, and $A>0$ after then. The finally used interplanetary parameters 
are shown in Fig.~\ref{fig:MF_TA_A}

\begin{figure}[!ht]
\centering
\includegraphics[width=\columnwidth]{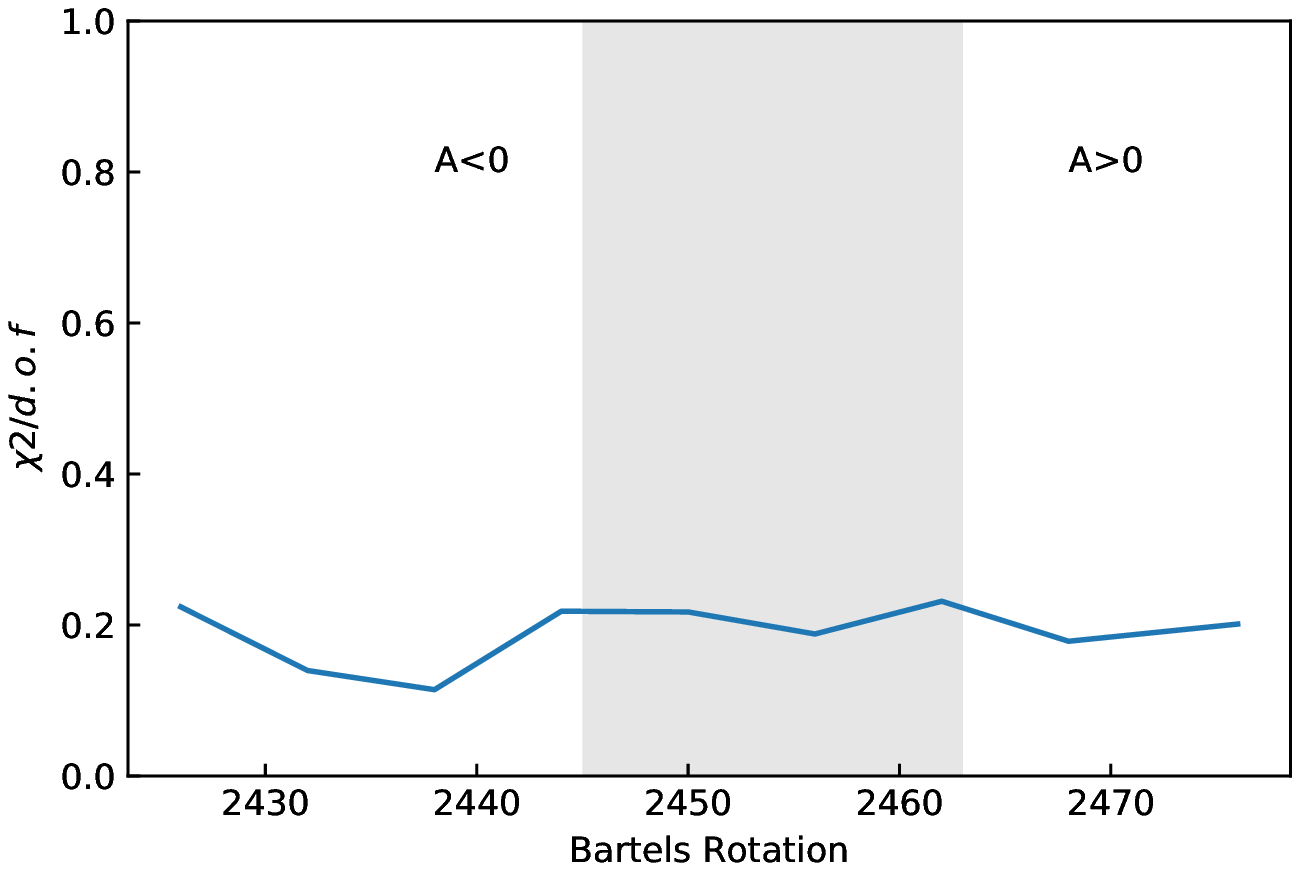}  
\caption{The reduced $\chi^2$ for the fittings of the time-dependent proton fluxes. 
The period with uncertain HMF polarity is marked in gray.} 
\label{fig:chi2}
\end{figure}

\begin{figure*}[!ht]
\centering
\includegraphics[width=\textwidth]{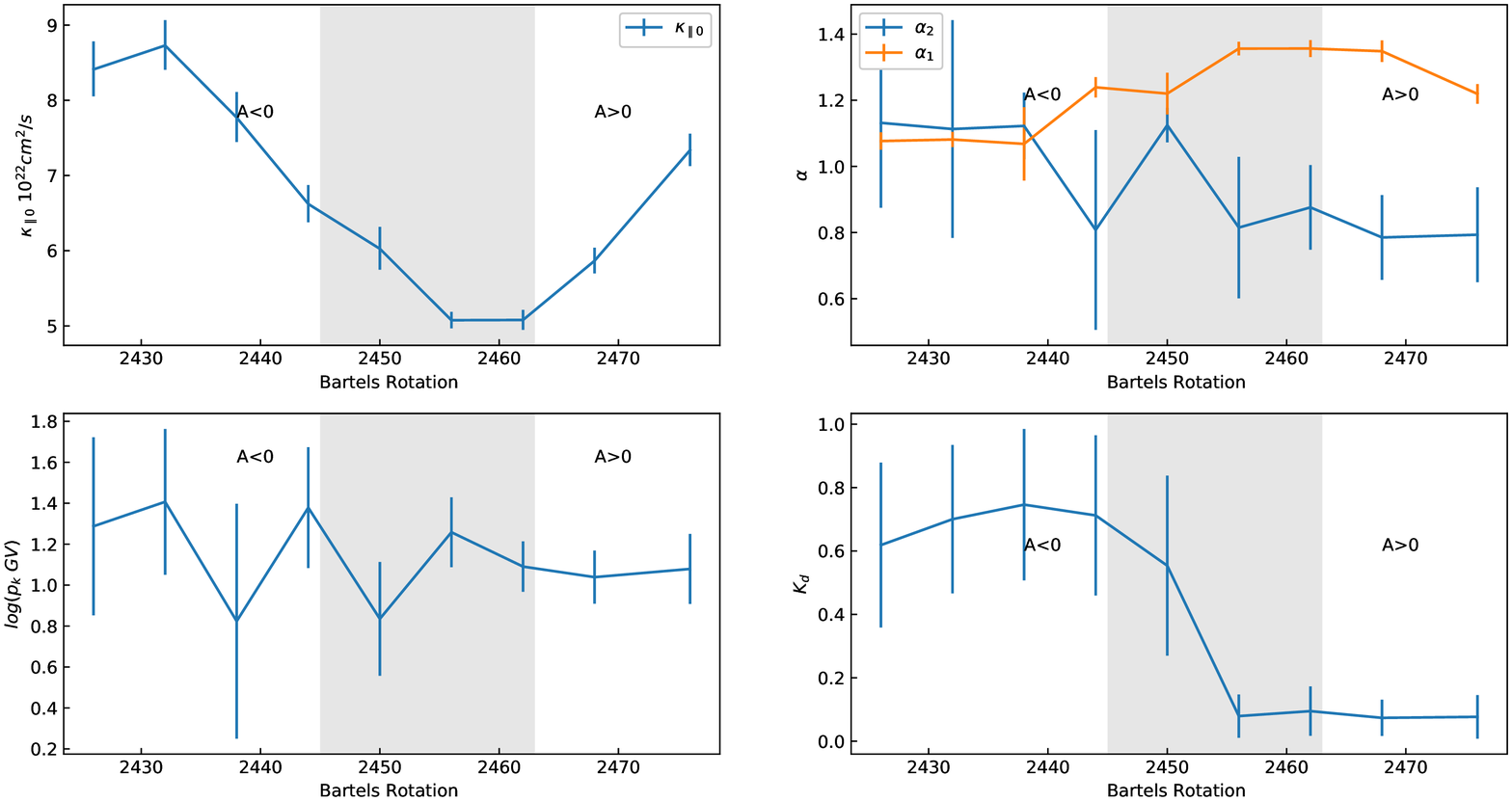}  
\caption{The fitting solar modulation parameters and $1\sigma$ uncertainties. 
The period with uncertain HMF polarity is marked in gray.
} 
\label{fig:param}
\end{figure*}

The reduce $\chi^2$ of the fittings is show in Fig.~\ref{fig:chi2}, and the best-fit 
parameters of $\kappa_{\parallel}^0$, $\alpha_1$, $\alpha_2$, $p_k$, and $K_d$ are shown 
in Fig.~\ref{fig:param}. We can see that all the parameters correlate with the polarity
reversal of the HMF, indicating the importance of considering the time-dependent
modulation effect in detail rather than an average modulation effect. 
The diffusion coefficient $\kappa_{\parallel}^0$ is inversely correlated with the
solar activities. This is understandable, since more turbulent of the interplanetary
environment means a slower diffusion. The drift suppression factor $K_{\rm d}$ is larger 
for $A<0$, and smaller for $A>0$, consistent with previous work \cite{Ferreira_2004,Song_2021}.
Our fittings further show that probably no break of the diffusion coefficient is
required when $A<0$, given the very close values of $\alpha_1$ and $\alpha_2$.

\subsection{Impact of uncertainties of the solar modulation fitting}

\begin{figure}[!ht]
\centering
\includegraphics[width=\columnwidth]{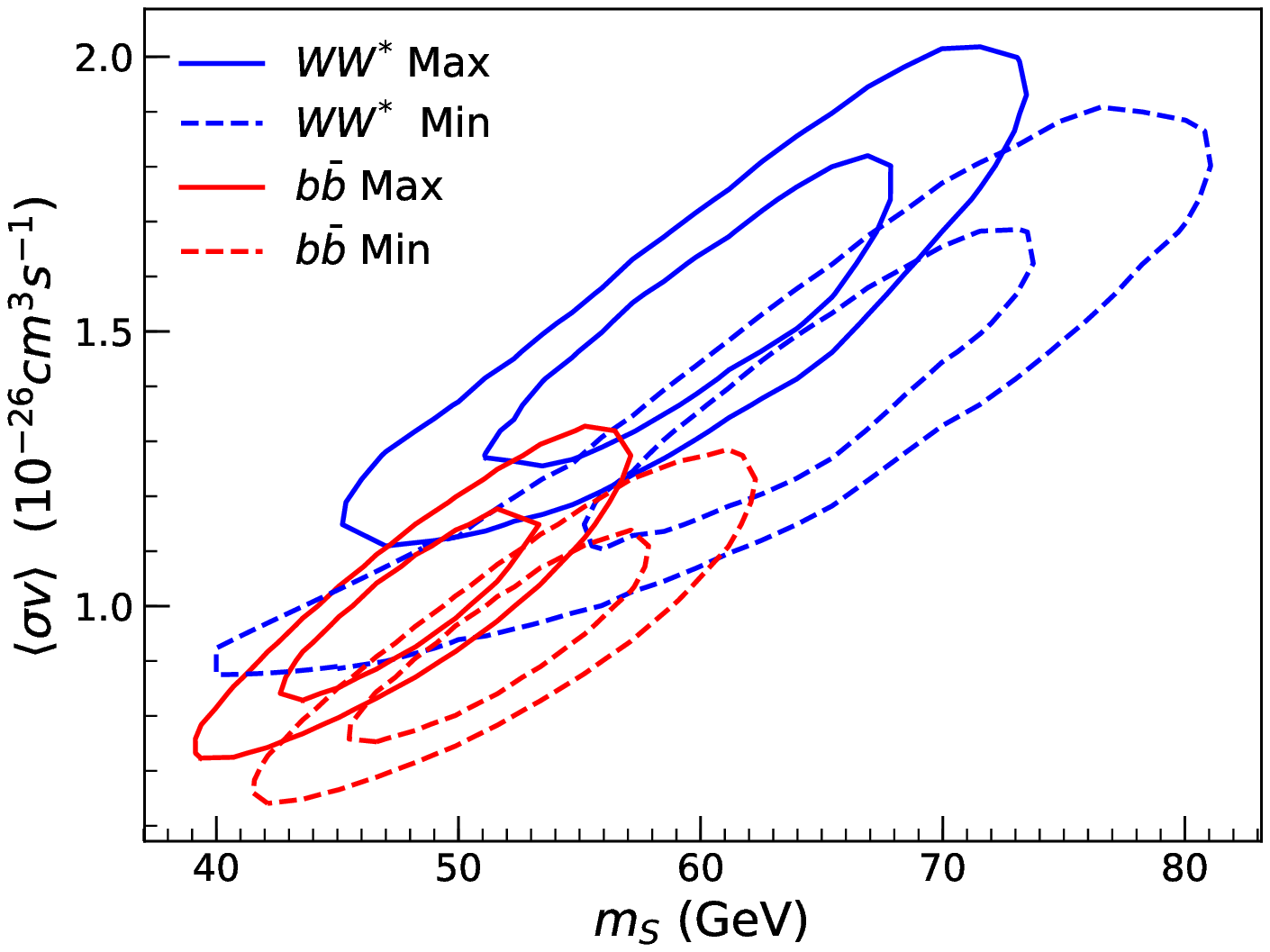}  
\caption{The favored DM parameter regions (1$\sigma$ and 2$\sigma$) by antiprotons 
for solar modulation parameters which enveloping the measurement uncertainties of protons.}
\label{fig:moduHL}
\end{figure}

To quantify the uncertainties of the solar modulation fitting, we adopt the modulation 
parameters which match the upper and lower boundaries of the proton data and repeat the 
likelihood fitting of the DM contribution to antiprotons. The resulting $1\sigma$ and 
$2\sigma$ constraints on the $(m_S,~\langle\sigma v\rangle)$ parameters are shown in 
Fig.~\ref{fig:moduHL}. This results in uncertainties of $\sim20\%$, mainly on the 
annihilation cross section.

\subsection{The ROI, the likelihood map and another spectral component of the GCE spectral template fitting}

\begin{figure*}
\centering 
\includegraphics[width=1\textwidth]{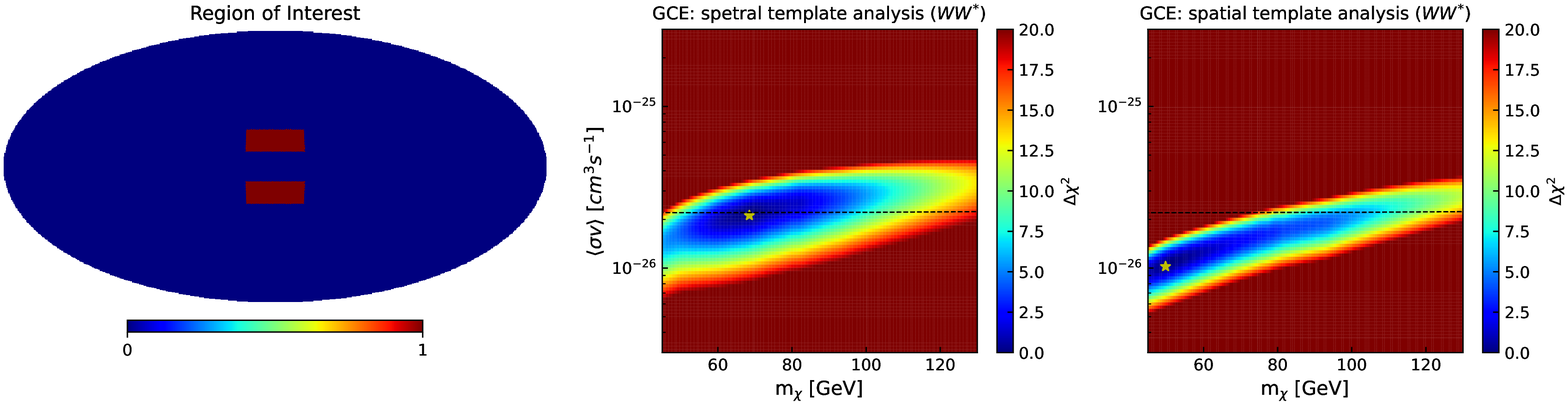}  
\caption{ Left: The ROI (brown) used in the GCE fitting. Middle: The $\Delta \chi^2$ map on the mass and cross section parameter plane for the $SS\to WW^*$ channel with the spectral template analysis. Right: The $\Delta \chi^2$ map with the spatial template analysis. The yellow stars show the best-fit DM parameters. The dashed black lines represent the mass-dependent relic annihilation cross section calculated in Ref.~\cite{Steigman2012}}
\label{fig:GCE}
\end{figure*}

The left panel of Fig.~\ref{fig:GCE} shows the ROI of our analysis. The middle and right panels show the likelihood maps, characterized by $\Delta \chi^2=\chi^2 - \chi^2_{\rm min}$\footnote{See the definition of $\chi^2$ in Ref.~\cite{Huang:2015rlu}}, on the 
($m_S$, $\left\langle \sigma v \right\rangle$) parameter plane for the $SS\to WW^*$ channel.

{\bf \subsection{The GCE spatial template analysis}}

\begin{figure}
\centering 
\includegraphics[width=\columnwidth]{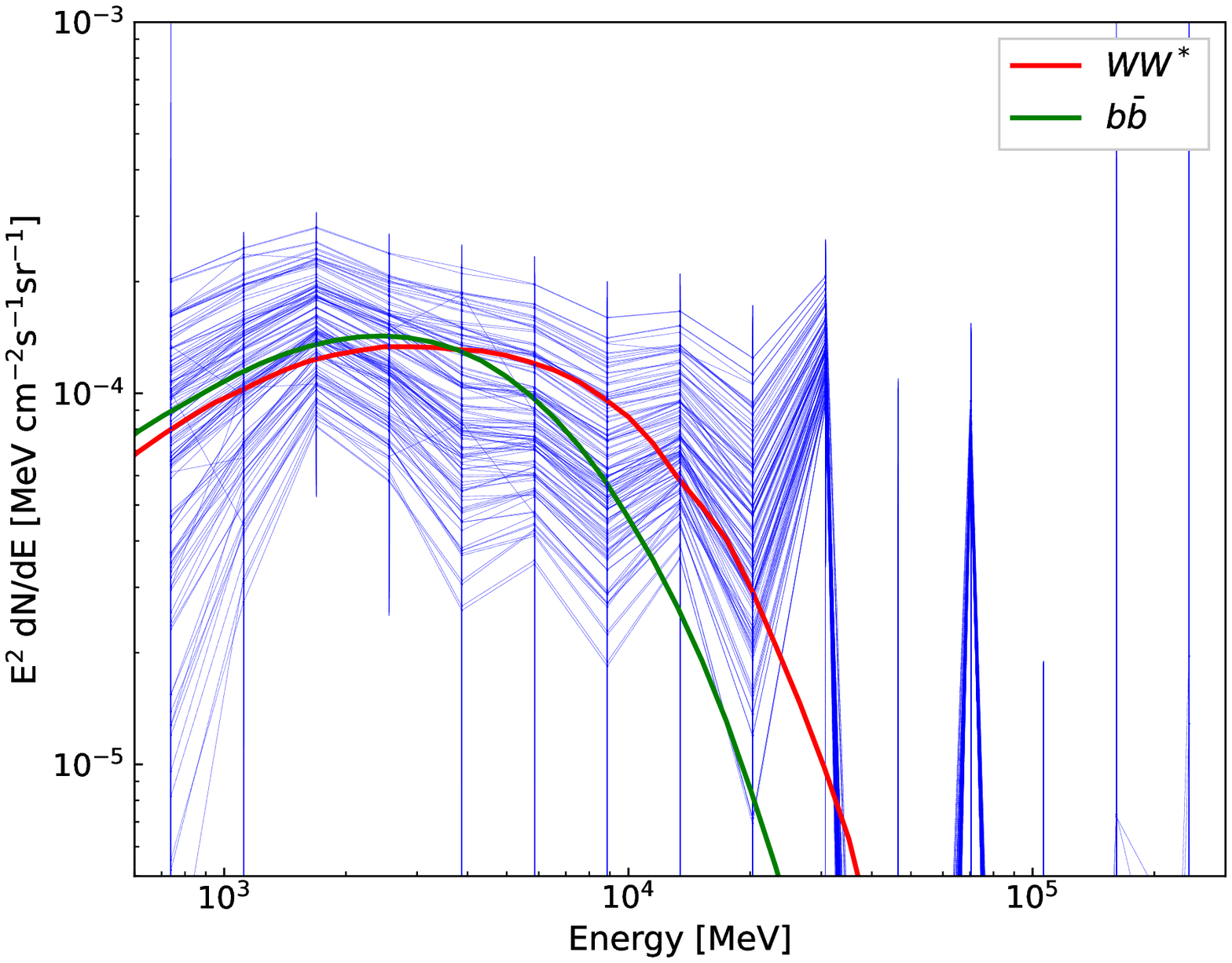}  
\caption{ The SED for the GCE component obtained from the spatial template analysis with 128 galactic diffuse emission templates. The red (green) line represents the spectrum with the best-fit DM parameters for the $WW^*$ ($b\bar{b}$) channel.}
\label{fig:GCE_ST_SED}
\end{figure}

{For the spatial template analysis, we adopt a similar method as in previous work \cite{Zhou:2014lva, Calore:2014xka, Cholis:2021rpp}. We use Fermi Pass 8 data, version P8R3 and class ULTRACLEANVETO, recorded from Aug 4 2008 to Feb 1 2020, in total 600 weeks. To suppress the contamination from $\gamma$-ray generated by CR interactions in the upper atmosphere, photons collected at zenith angles larger than 90$^\circ$ are removed. Moreover, we adopt the specifications (DATA$\_$QUAL$>$0) $\&\&$(LAT$\_$CONFIG==1) to select the good quality data. We use the healpix projection \cite{Gorski:2004by} for the spatial binning of data, with the  resolution parameter nside = 256. And we also bin the data, from 600 MeV to 300 GeV, into 15
logarithmically distributed energy bins. We use the Fermitools version 1.2.1 to calculate the relevant exposure-cube files and exposure maps, with the instrumental response function P8R3\_ULTRACLEANVETO\_V2.}

{The likelihood for estimating the contribution from different components is constructed as 
\begin{equation}
-2 \ln \mathcal{L}=2 \sum_{i, j} \left(\mu_{i, j}-k_{i, j} \ln \mu_{i, j}\right)+\chi_{\mathrm{ext}}^2,
\end{equation}
where $\mu_{i,j}$ and $k_{i,j}$ are the expected and observed number of photons in the energy bin $i$ and the pixel $j$, respectively, and $\chi_{\rm ext}^2$ is used to further constrain the contribution from some model components. For the expected number of photons, we take}

{\begin{equation}
\begin{aligned} \mu_{i,j} &=q_{i}^{\pi^0+bremss}N^{\pi^0+bremss}_{i,j} +q_{i}^{ics}N^{ics}_{i,j}+q_{i}^{iso}N^{iso}_{i,j}\\&+q_{i}^{bubble}N^{bubble}_{i,j}+q_{i}^{psc}N^{psc}_{i,j}
\\&+q_{i}^{gNFW}N^{gNFW}_{i,j}+q_{i}^{\overline{gNFW}}N^{\overline{gNFW}}_{i,j}
, \end{aligned}
\end{equation}}
{ where $N^{\pi^0+bremss}_{i,j}$, $N^{ics}_{i,j}$, $N^{iso}_{i,j}$, $N^{bubble}_{i,j}$, $N^{psc}_{i,j}$, $N^{gNFW}_{i,j}$, and $N^{\overline{gNFW}}_{i,j}$ are predicted number of photons for the galactic diffuse $\pi^0$ decay emissions plus the  bremsstrahlung emissions, the inverse Compton scattering emissions, the  isotropic $\gamma$-ray emissions, the Fermi Bubbles, point sources, the GCE in the region defined in the left panel of Fig.~\ref{fig:GCE}, and the "GCE-like" component in the  complement of the above mentioned region, respectively.   
And q$_{i}$ stands for the scaling factor for each component in each energy bin. For the galactic diffuse emissions, we use 128 templates introduced in Ref.~\cite{Fermi-LAT:2012edv}, and we tie the $\pi^0$ decay emissions and the  bremsstrahlung emissions into a single component since morphologies of these two emissions are both primarily determined by the interstellar medium gas \cite{Calore:2014xka,Cholis:2021rpp}. For the isotropic $\gamma$-ray emissions, we use a flat and homogeneous flux to take into account the combined emission from unresolved point sources and misidentified cosmic rays. For the Fermi Bubbles, we adopt the spatial template from Ref.~\cite{Su:2010qj}. For point sources, the predicted photon counts are calculated by {\it gtmodel} for all detected sources in the 4FGL catalog \cite{Fermi-LAT:2019yla,Fermi-LAT:2022byn}. For the GCE in the interested region, we take the same generalized Navarro-Frenk-White profile as in the antiproton and spectral template analysis. And for the "GCE-like" component, we use the same generalized Navarro-Frenk-White profile but use an independent factor to take into account the possible contribution from unresolved sources in the bulge \cite{Macias:2016nev,Bartels:2017vsx}. We 
multiply these templates with exposure maps to get the predicted photon counts. And we also use the Fermi-LAT PSF calculated by {\it gtpsf} at the galactic center position to smooth all components to account for the finite angular resolution of the instrument.}

{We first choose the region satisfying $\left| b\right| \textless -2^{\circ}$ in the southern sky with less complicated structures, and fit the likelihood with $\chi^2_{\rm ext}$ set to zero. And we could get the estimation about the fluxes and statistical errors for the isotropic $\gamma$-ray emissions and the Fermi Bubbles, respectively. Then we choose the inner galaxy region, $|l| \leq 20^{\circ}$  and $ 2^{\circ} \leq|b| \leq 20^{\circ}$, and fit the likelihood with constraints for  the isotropic $\gamma$-ray emissions and the Fermi Bubbles. The constraints are assumed to be Gaussian, as
\begin{equation}
\chi_{\mathrm{ext}}^2=\sum_{i, k}\left(\frac{\phi_{i, k}-\bar{\phi}_{i, k}}{\Delta \phi_{i, k}}\right)^2
\end{equation}
where $\phi_{i,k}$ is the predicted flux of component $k$ in energy bin $i$, $\bar{\phi}_{i,k}$ are the estimated fluxes from the $\left| b\right| \textless -2^{\circ}$ region and $\Delta \phi_{i,k}^2$ are the sum of statistical and systematic errors from the effective area, based on analysis in the $\left| b\right| \textless -2^{\circ}$ region too, in quadrature. Then we can get the scaling factor q$_{i}$ for each component in each energy bin, and get the spectral energy distribution (SED) for the GCE component. 
To account for the systermatics from the background modeling, we repeat above analysis for total 128 galactic diffuse emission templates and derive SED for the GCE component as shown in Fig.~\ref{fig:GCE_ST_SED}. }

{Then we scan the DM parameters ($m_S$, $\left\langle \sigma v \right\rangle$) to calculate the $\chi^2$ map for each SED with
\begin{equation}
\chi^2=\sum_{i}\left(\frac{\phi_{i}(m_S,\left\langle \sigma v \right\rangle)-\bar{\phi}_{i}}{\Delta \phi_{i}}\right)^2,
\end{equation}
where $\phi_{i}(m_S,\left\langle \sigma v \right\rangle)$ is the predicted flux from the DM annihilation at the GCE region in the $i$ th energy bin, $\bar{\phi}_{i}$ is the measured flux from the SED we obtained and $\Delta \phi_{i}^2$ is the orthogonal sum of corresponding statistical and systematic errors from the effective area.
The final result of the spatial template analysis is derived from the average of all 128 $\chi^2$ maps.
The averaged $\Delta \chi^2$ (defined as $\chi^2 - \chi^2_{\rm min}$) map is shown in the right panel of Fig.~\ref{fig:GCE}.}

{\bf \subsection{The result for the $b\bar{b}$ channel}}

 \begin{figure}[!ht]
 \centering
 \includegraphics[width=\columnwidth]{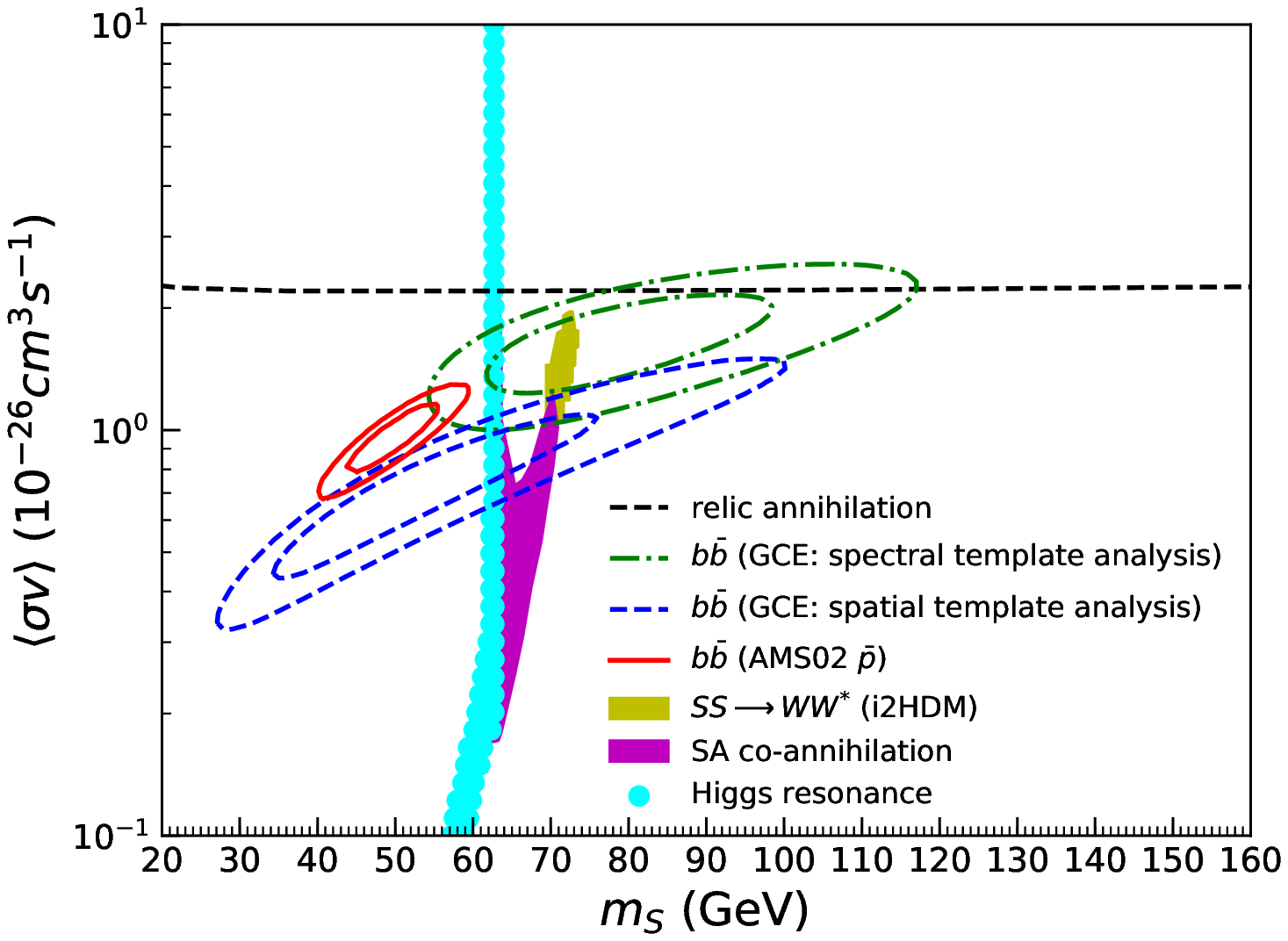}  
 \caption{The favored DM parameter space via fitting to the antiproton and GCE data (1$\sigma$ and 2$\sigma$ from inside to outside) for the $b\bar{b}$ channel, as well as the i2HDM model parameters to fit the $W$-boson mass anomaly (the 95\% region, adopted from Ref.~\cite{i2HDM}).
The black dashed line is the mass-dependent relic annihilation cross section \cite{Steigman2012}.}
 \label{fig:bbar}
 \end{figure}

{Fig.~\ref{fig:bbar} shows results from the antiproton and GCE analysis for the channel $SS\to b\bar{b}$.}

\clearpage 

\bibliographystyle{apsrev}
\bibliography{refs}

\end{document}